\documentclass[osajnl,amsmath,amssymb,twocolumn,showpacs,superscriptaddress]{revtex4}%
\usepackage{amsmath}
\usepackage{xcolor}
\usepackage{graphicx}
\usepackage{dcolumn}
\usepackage{bm}
\usepackage{float}
\usepackage{epsfig}
\usepackage{graphicx}%
\usepackage{amsmath}%
\usepackage{amsfonts}%
\usepackage{amssymb}
\usepackage{natbib}
\usepackage{soul}
\usepackage{color}

\usepackage{amssymb}
\usepackage{graphicx}
\begin{document}
\title{Cavity solitons in a microring dimer with gain and loss}

\author{Carles Mili\'{a}n}
\email{carles.milian@icfo.eu}
\affiliation{ICFO--Institut de Ciencies Fotoniques, The Barcelona Institute of Science and Technology, 08860 Castelldefels (Barcelona), Spain}
\author{Yaroslav V. Kartashov}
\affiliation{ICFO--Institut de Ciencies Fotoniques, The Barcelona Institute of Science and Technology, 08860 Castelldefels (Barcelona), Spain}
\affiliation{Institute of Spectroscopy, Russian Academy of Sciences, Troitsk, Moscow Region, 142190, Russia}
\affiliation{Department of Physics, University of Bath, Bath BA2 7AY, UK}
\author{Dmitry V. Skryabin}
\affiliation{Department of Physics, University of Bath, Bath BA2 7AY, UK}
\affiliation{Department of Nanophotonics and Metamaterials, ITMO University, St. Petersburg 197101, Russia}
\author{Lluis Torner}
\affiliation{ICFO--Institut de Ciencies Fotoniques, The Barcelona Institute of Science and Technology, 08860 Castelldefels (Barcelona), Spain}
\affiliation{Universitat Polit\`{e}cnica de Catalunya, 08034 Barcelona, Spain}
%
%
%
%
%
\begin{abstract}
We address a pair of vertically coupled microring resonators with gain and loss pumped by a single-frequency field. Coupling between microrings results in a twofold splitting of the single microring resonance that increases when gain and losses decrease and that gives rise to two different cavity soliton (CS) families. We show that the existence regions of CSs are tunable and that both CS families can be stable in the presence of an imbalance between gain and losses in the two microrings. These findings enable experimental realization of frequency combs in configurations with active microrings and contribute towards the realization of compact multisoliton comb sources.
\end{abstract}
\maketitle
%
%
Nowadays microring resonators are among the most widely used devices for generation of optical frequency combs \citep{kippenberg2011microresonator}. Particularly important is the ability to tune combs by modifying the free spectral range (FSR) of the resonator \cite{savchenkovPRL2008} or by impinging a global frequency shift over a whole FSR via, e.g., thermal effects \cite{delHayePRL2011}. Cavity solitons (CSs) play a central role in the formation of coherent frequency combs because they remain robust under strong higher-order linear \cite{tlidiOL10,tlidiPRA13,erk,parra,milianOE,brasch2016,mbe2017} and nonlinear \cite{milianPRA,karpov2016raman,VAHALA} effects. CSs in focusing media appear in the red-detuned region around the cavity resonance closest to the pump frequency, while their observation in the blue-detuned region is possible only in special cases, such as arising with a bi-chromatic pump \cite{HanssonPRA2014}.

Especially rich possibilities for frequency comb tuning and light switching appear in geometries involving dimers and arrays of microring resonators. Importantly, splitting of resonances due to coupling between several microrings \cite{Smith03} and their overlap due to nonlinearity leads to multistability \cite{Dum2005TRISTABILITY} with a much lower phase shift threshold than the multistability achieved by the overlap of single ring resonances \cite{Dum2005TRISTABILITY}. While multistability in single microrings \cite{hansson2015frequency,And2017,Kartashov17,conforti2017multi} and birefringent fiber loops \cite{tlidiOL17} has attracted considerable interest for super-cavity soliton formation, the first results on comb tunability \cite{miller2015tunable} and soliton formation \cite{Kipp_stack2016} in coupled microrings were obtained only recently and they are expected to boost CS-based applications. Introduction of active elements into system of coupled microrings opens new prospects in frequency comb generation. Thus, dimers consisting of passive and active microrings are under very active investigation in the context of $\mathcal{PT}$-symmetry. Such effects as optical isolation \cite{chang2014}, unidirectional transmission \cite{PengNPhys14}, lasing revival \cite{PengScience14}, 
coherent perfect absorption \cite{longhi2014pt}, selective single mode lasing \cite{feng2014,hod2014},
 and reversal of the $\mathcal{PT}$ phase transition \cite{hassan2015}, have been demonstrated. Dimers with nonlinear losses also exhibit rich nonlinear dynamics \cite{konotop2017}. Nevertheless, despite the considerable amount of work reported on solitons in $\mathcal{PT}$-symmetric systems \citep{konotopRMP16,suchkov2016}, they have not been addressed in $\mathcal{PT}$-symmetric microrings with driving. Solitons in active cavities have been studied in VCSELs with frequency selective feedback in one and two dimensions \cite{paulau07,paulau08,paulau10,paulau11}, where they are described by different models employing dissipative nonlinearities or coupling.

\begin{figure}
\begin{center}
\includegraphics[width=.49\textwidth]{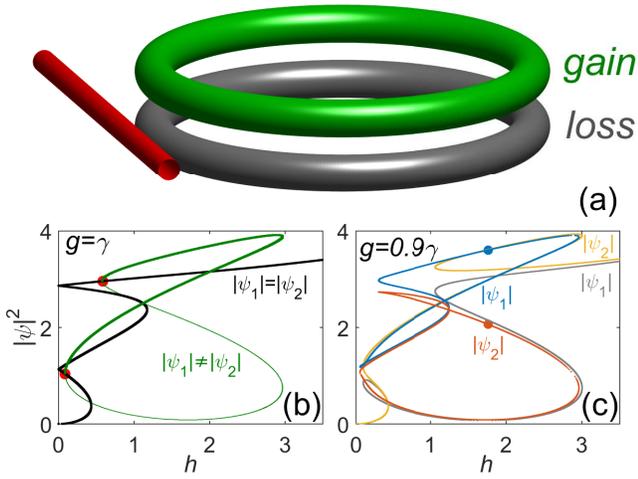}
\caption{(a) Sketch of the externally driven microring dimer. The pump waveguide, equidistant from both rings, provides equal driving strengths. Amplitudes $|\psi_{1,2}|$ versus $h$ in the $\mathcal{PT}$-symmetric case $g=\gamma$ (b), and with gain/loss $g=0.9\gamma$ (c) for $\delta=2$ and $\gamma=0.5$. Dots in (b) mark the local bifurcations where asymmetric states (green lines) coalesce with symmetric ones (black lines). Dots in (c) mark a state $\{|\psi_1|,|\psi_2|\}$. \label{f1}}
\end{center}
\end{figure}

In this Letter we address driven microring dimers with gain and loss, as sketched in Fig. \ref{f1}(a), and show that they support two different  CS families that are stable for different cavity detunings and which stability domains are found to be tunable with gain and losses. We have chosen a vertical coupling geometry because it is advantageous for designing compact multi-soliton sources \cite{Kipp_stack2016}. In such a structure, resonance splitting (yielding multistability in the nonlinear regime) is determined by the interplay between coupling and gain/loss, providing new opportunities for tuning the comb. Our work extends studies of solitons in $\mathcal{PT}$-symmetric couplers \cite{BarashPRA2013,DribenOL2011} to externally driven systems. Moreover, external driving allows obtaining stable solitons even in the presence of imbalance between gain and loss.

We study the light evolution in the microring dimer governed by the coupled Lugiato-Lefever equations \citep{lugiato1987,chembo2010}:
\begin{eqnarray}
\nonumber &&
-i\partial_t\psi_1=\frac{B_2}{2}\partial_x^2\psi_1+(i\gamma-\delta+|\psi_1|^2)\psi_1+\kappa\psi_2+h
,\\ &&
-i\partial_t\psi_2=\frac{B_2}{2}\partial_x^2\psi_2+(-ig-\delta+|\psi_2|^2)\psi_2+\kappa\psi_1+h
\label{eq:model2},
\end{eqnarray}
where $\gamma>0$ and $g>0$ account, respectively, for loss and gain in rings 1 and 2; $\delta\equiv[\omega_0-\omega_p]\tau$ is the normalized cavity detuning; $\omega_p$ is the pump frequency; $\omega_0$ is the closest resonance frequency of the cavity; $\tau\equiv2\pi Rn_g/c$ is the roundtrip time for the pump frequency; $n_g$ is the group index at $\omega_{p}$; $Q=\omega_p\tau/\gamma$ is the quality factor; $R$ is the microring radius; $B_2=\omega^{(2)}\tau/(2\pi R)^2$ is the group velocity dispersion coefficient; $\omega^{(2)}\equiv\partial_\beta^2\omega\vert_{\beta_0}$; $\beta$ is the propagation constant; $\psi=E\sqrt{{n_{NL}}\tau}$ and $h=\tau\sqrt{\tau{n_{NL}}}\omega_p^2\mathcal{S}/\omega_0$, where $E$, ${n_{NL}}$, $\mathcal{S}$ are the physical field, nonlinear coefficient, and coupled pump strength. Coupling between rings $\kappa$$=\omega_0\tau\mathcal{I}/2$, where $\mathcal{I}$ is the overlap integral of modes in rings 1 and 2. Equations \ref{eq:model2} assume coupling via evanescent mode tails, hence it is purely linear \citep{konotopRMP16,suchkov2016}. $t$, $x$ are the normalized time, coordinate along the microring, respectively. Eq. \ref{eq:model2} is invariant under the scaling $\{t,x,\psi_{1,2},\gamma,g,\delta,\kappa,h\}\rightarrow\{ta,x\sqrt{a},\psi_{1,2}/\sqrt{a},\gamma/a,g/a,\delta/a,\kappa/a,ha^{-3/2}\}$. Below we set $a=\kappa$ and also make use of the transformation $x\rightarrow x\sqrt{B_2}$. The above results in recasting Eq.\ref{eq:model2} as if we had taken $B_2=\kappa=1$.

\begin{figure}
\centering
\includegraphics[width=.49\textwidth]{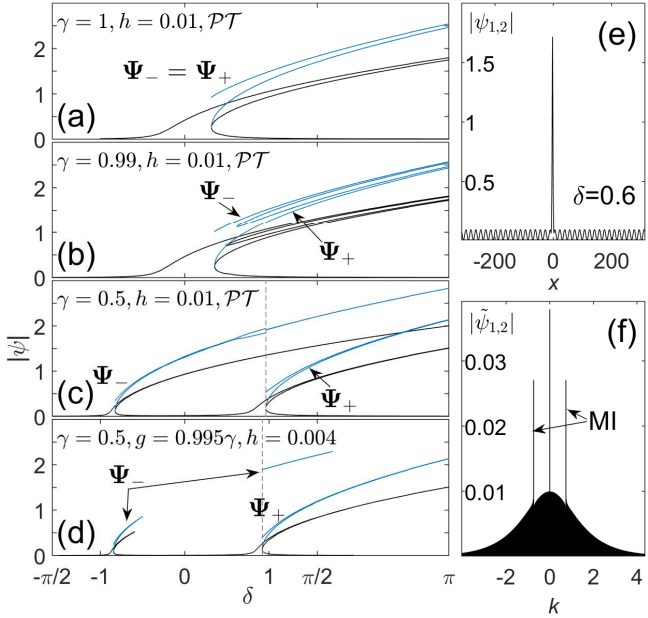}
\caption{Nonlinear resonances vs detuning (black lines). (a,b,c) $\mathcal{PT}$-symmetric case with $\gamma=1$, $0.99$, $0.5$. (d) broken $\mathcal{PT}$-symmetry case with $g/\gamma=0.995$. Soliton branches are shown for the $\boldsymbol{\Psi}_\pm$ families (blue lines). (e,f) spatial and spectral profiles of a $\boldsymbol\Psi_-$ soliton at $\delta=0.6$ from the top branch of (c).
\label{fsplit}}
\end{figure}

Homogeneous solutions ($\partial_t\psi=\partial_x\psi=0$) of Eq. \ref{eq:model2} are found from the equations: $\psi_1=
h[1+\delta-|\psi_2|^2+ig]/\Delta$, $\psi_2=
h[1+\delta-|\psi_1|^2-i\gamma]/\Delta$, and $\Delta(|\psi_{1,2}|)\equiv(i\gamma-\delta+|\psi_1|^2)(-ig-\delta+|\psi_2|^2)-1$. At $|g|=|\gamma|$, $|\psi_{1,2}|=|\psi|$ these equations reduce to
\begin{eqnarray}
h^2=\left(|\psi|^3-(\delta-1)|\psi|\right)^2+\gamma^2|\psi|^2-\frac{2\gamma(\gamma+g)|\psi|^2}{[|\psi|^2-(1+\delta)]^2+\gamma^2}
\label{eq:hmultistab}.
\end{eqnarray}
If the two rings are passive and identical, $g=-\gamma$, one obtains a standard bi-cubic equation for the field amplitude, with effective detuning $\delta-1$, describing known bistability \cite{BarashPRE1996}. In the $\mathcal{PT}$-symmetric case ($g=\gamma$) Eq. \ref{eq:hmultistab} has five real roots when $\gamma<1$ and $\delta>(1-\gamma^2)^{1/2}$. Numerically calculated dependencies $|\psi(h)|$ and $|\psi(\delta)|$ illustrating multistability are shown in Figs. \ref{f1}(b,c) and Figs. \ref{fsplit}(a-d), respectively. Figure \ref{f1}(b) shows $|\psi(h)|$ dependence in the  $\mathcal{PT}$-symmetric case. Solutions with $|\psi_{1}|=|\psi_{2}|$ emerge from the origin (see black line), but in addition to them asymmetric states are born at local bifurcations (red dots) and exist only for $h>0$. When $|\psi_1|$ emerges from a bifurcation point on the thick green branch, $|\psi_2|$ follows the thin green line, and vice versa. Figure \ref{f1}(c) illustrates how the imbalance between gain and loss breaks degeneracy \cite{hassan2015}.

Figures \ref{fsplit}(a-d) with $|\psi(\delta)|$ curves (black) illustrate that the multistability originates from resonance splitting mediated by coupling between the two microrings. Indeed, in the low-pump limit flat solutions $|\psi|^2_{h=0}=\delta\pm\sqrt{1-\gamma^2}$ show that two resonances emerge at $\delta=\pm\sqrt{1-\gamma^2}$. These resonances tilt in the nonlinear case leading to multistability that is controlled by  coupling and $g$,$\gamma$: at $\gamma=1$ the resonances merge into one (Fig. \ref{fsplit}(a)) and for $\gamma<1$ the resonances shift away from each other (Figs. \ref{fsplit}(b,c)). For $g<\gamma$ the resonances shorten and multistability may disappears (Fig. \ref{fsplit}(d)). Splitting of resonances in our system occurs for $\gamma<1$, i.e., for $Q>4\pi^2Rn_g/(\lambda_p\kappa)$. Validity of Eq. \ref{eq:model2} requires $\kappa\ll1$. Hence, considering $\kappa\approx0.1$, observation of the resonance splitting in a silica glass or silicon nitride microring with $R\sim100\ \mu$m at $\lambda_p=1.5\ \mu$m ($n_g\sim1.5$, $\tau\sim3$ ps, $B_2\sim10^{-5}$) requires $Q\gtrsim6\times10^4$. This is easily accomplished as usually $Q\gtrsim10^6$, and hence balance condition $g=\gamma$ is met for gain rates $g\kappa/\tau\leq\omega_p/Q\approx1$ GHz, well located within the intervals of gain experimentally attainable \cite{chang2014}. External driving power required for single soliton excitation with the above parameters is $\sim1$ W \cite{karpov2016raman}, which roughly corresponds to $h=0.001$. With the above, $FSR\approx400$ GHz and $\delta=1$ corresponds to $\sim30$ GHz.
%
\begin{figure}
\centering
\includegraphics[width=.49\textwidth]{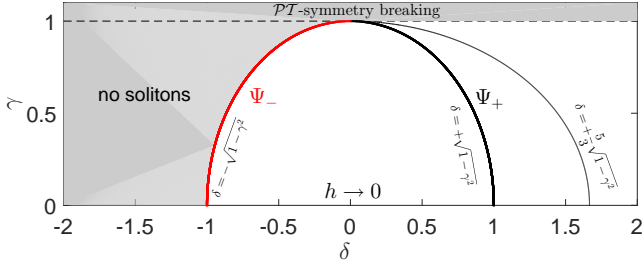}
\caption{Frequency-locked soliton families existence threshold (solid curves) and stability limit (dashed curve) in the low pump and $\mathcal{PT}$-symmetric limits. While $\boldsymbol{\Psi_{-}}$ are always unstable, $\boldsymbol{\Psi_{+}}$ are stable within a finite range of detuning. \label{fsol1}}
\end{figure}

We now turn to localized solutions of Eq. \ref{eq:model2}. In the limit $h\rightarrow0$ and $g=\gamma$ two soliton families can be found analytically:
\begin{eqnarray}
\psi_{1,\pm}=\eta_\pm f_\pm^{-1/2}\mathrm{Sech}(\eta_\pm x),\label{eq:couplersol1}\ \ \psi_{2,\pm}=f_\pm\psi_{1,\pm}=\psi_{1,\pm}^*,\\ 
\eta_\pm^2=2\delta\mp2(1-\gamma^2)^{1/2}\in\mathrm{Re},\ 
f_\pm=\pm \exp(\mp i\arcsin(\gamma)).
\label{eq:couplersol}
\end{eqnarray}
We denote these two families by ${\boldsymbol{\Psi_\pm}}\equiv\{\psi_{1,\pm},\psi_{2,\pm}\}$. 
Figure \ref{fsol1} shows existence and stability chart for the solitons in the $\mathcal{PT}$-symmetric system in the low-pump limit. $\boldsymbol{\Psi_{\pm}}$ families exist above the threshold detuning $\delta=\pm(1-\gamma^2)^{1/2}$ (see Eq. \ref{eq:couplersol}). While the $\boldsymbol{\Psi}_-$ family is always unstable for $h\rightarrow0$ \cite{BarashPRA2013}, $\boldsymbol{\Psi}_+$ family is stable for $\eta^2\leq(4/3)(1-\gamma^2)^{1/2}$ \cite{DribenOL2011,BarashPRA2013} which, together with Eq. \ref{eq:couplersol}, yields stability at $(1-\gamma^2)^{1/2}\leq\delta\leq(5/3)(1-\gamma^2)^{1/2}$.

In what follows we discuss solitons in the microring dimer under external driving. Equations \ref{eq:model2} predict that power integrals $\mathcal{P}_{1,2}\equiv\int|\psi_{1,2}|^2\mathrm{d}x$ vary in accordance with
\begin{eqnarray} \partial_t(\mathcal{P}_1+\mathcal{P}_2)=-2\Gamma(\mathcal{P}_1+\mathcal{P}_2)+2h\int\left( \mathrm{Im}(\psi_1)+\mathrm{Im}(\psi_2)\right)\mathrm{d}x\label{eq:pow},
\end{eqnarray}
where integrals span over the microring circumference and $\Gamma\equiv[\gamma\mathcal{P}_1-g\mathcal{P}_2]/[\mathcal{P}_1+\mathcal{P}_2]$ is the loss/gain coefficient of the dimer (note $\Gamma=0$ for $\mathcal{PT}$-symmetric states with $\mathcal{P}_1=\mathcal{P}_2$).
Importantly, the presence of driving in Eq. \ref{eq:pow} suggests that stationary solutions of Eq.\ref{eq:model2} exist not only for $g=\gamma$ 
, but also for $g\neq\gamma$
, contrarily to the case with $h=0$ \citep{BarashPRA2013}. Below we show that the two soliton families $\boldsymbol{\Psi_\pm}$ from the unforced case (Eqns. \ref{eq:couplersol1} and \ref{eq:couplersol}) are continuously connected to solitons of the driven dimer. We anticipate that for $h>0$ stable $\boldsymbol{\Psi_+}$ solitons exist under ideal and broken $\mathcal{PT}$-symmetry (stability was tested by performing rigorous linear stability analysis). On the contrary, stable $\boldsymbol{\Psi_-}$ solitons are found only for $g<\gamma$. Solitons above the $\mathcal{PT}$-symmetry breaking point, $\gamma>1$, were found to be unstable.

%
\begin{figure}
\centering
\includegraphics[width=.49\textwidth]{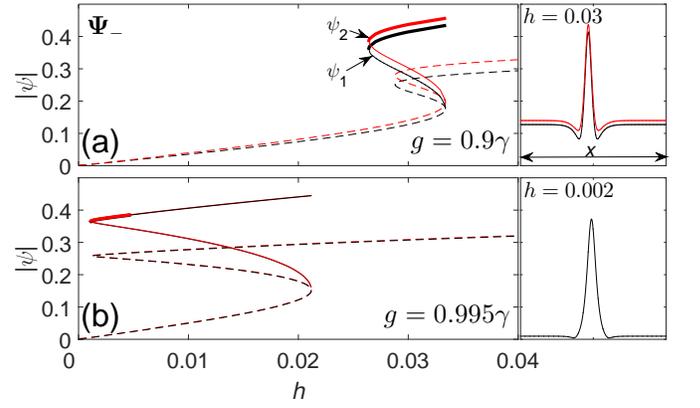}
\caption{$|\psi(h)|_{max}$ dependencies (solid) for $\boldsymbol{\Psi_-}$ soliton family with stability domains (solid thick) for $\delta=-0.8$ and $\gamma=0.5$: (a) $g=0.9\gamma$, (b) $g=0.995\gamma$. Dashed curves show amplitude of the background. Insets show selected soliton profiles $|\psi_{1,2}(x)|$. \label{psimin}}
\end{figure}
\begin{figure}
\centering
\includegraphics[width=.49\textwidth]{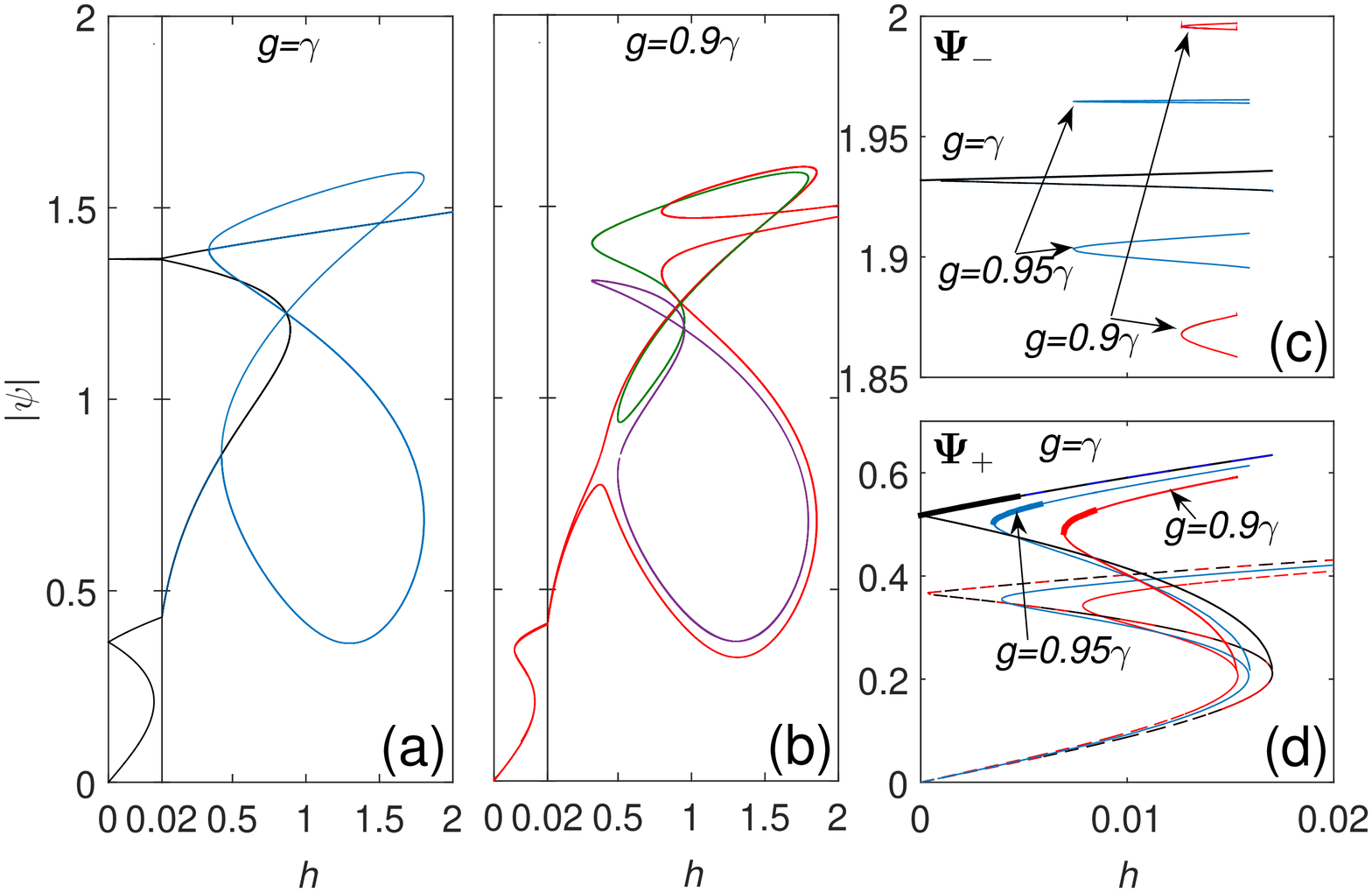}
\caption{(a,b) Amplitudes of flat solutions and (c,d)and solitons versus $h$ (solid thin curves) for the $\boldsymbol{\Psi_-}$ (c) and $\boldsymbol{\Psi_+}$ (d) families at $g/\gamma=0.9$, $0.95$, $1$ and $\delta=1$, $\gamma=0.5$. Thick curves denote soliton stability. Dashed curves in (d) background amplitude. \label{psiplus}}
\end{figure}
Solitons of the driven dimer are computed from Eq. \ref{eq:model2} with $\partial_t\psi_{1,2}=0$. Figures \ref{fsplit}(a-d) show soliton branches (i.e., soliton peak amplitude) as a function of detuning in different cases (see blue curves). When the resonances merge into one [$\gamma=1$, Fig. \ref{fsplit}(a)] the two soliton families are degenerate. This degeneracy is broken for $\gamma<1$. When $g$, $\gamma$ decrease, the detuning values at which $\boldsymbol{\Psi_\pm}$ families emerge gradually separate. These detuning values exactly coincide with the onset of bistability for each resonance. As expected, when pump tends to zero the width of the nonlinear resonances $|\psi(\delta)|$ decreases to zero and bistability thresholds in $\delta$ coincide with the solid curves in Fig. \ref{fsol1}.  

An unusual feature of the microring dimer is related to the $\boldsymbol\Psi_-$ family. Figure \ref{fsplit}(c) shows $\boldsymbol\Psi_-$ solitons emerging from the resonance at $\delta\approx-0.8$ and residing on the background corresponding to lower branch of left bistable loop. As $\delta$ is tuned towards the second resonance this background develops modulational instability (MI) and instability increases with $\delta$. Interestingly, despite instability of the background the system supports solitons with oscillatory tails. Figures \ref{fsplit}(e,f) show a soliton sitting on the modulated background, where MI bands are very narrow. Oscillations in soliton tails appear due to increasing overlap of the soliton spectrum with the MI bands and their amplitude is set by the exact balance between pump and frequency conversion processes. When $\delta$ crosses the multistability threshold, $\delta\approx0.96$, two different branches of $\boldsymbol\Psi_-$ exist. One is the continuation of the solitons  on the unstable background (not shown), and the other is the set of $\boldsymbol\Psi_-$ solitons sitting on the lowest and stable background supporting also the $\boldsymbol\Psi_+$ solitons. When small imbalance between gain and loss is introduced (Fig. \ref{fsplit}(d)), the leftmost resonance shortens and the two $\boldsymbol{\Psi_-}$ soliton branches become well separated in $\delta$. Despite the coexistence of the $\boldsymbol{\Psi_\pm}$ families at $\delta>0$, the $\boldsymbol{\Psi_-}$ family is found to be unstable for $\delta>0$, hence no stable coexistence (as in Refs. \cite{hansson2015frequency,And2017,Kartashov17,conforti2017multi,tlidiOL17}) was observed to occur.
%

Figures \ref{psimin} and \ref{psiplus} show the existence and stability domains for the $\boldsymbol{\Psi_{\pm}}$ families vs pump. In figure \ref{psimin} $\delta=-0.8$ and therefore only $\boldsymbol{\Psi_{-}}$ solitons exist (notice qualitative difference from these dependencies in VCSELs  \cite{paulau07,paulau08,paulau10,paulau11}). By decreasing the loss-gain imbalance from $g/\gamma=0.9$ (Fig. \ref{psimin}(a)) to $g/\gamma=0.995$ (Fig. \ref{psimin}(b)), bistability domain increases substantially but this is accompanied by a reduction of the soliton stability region, which vanishes completely for $g/\gamma\approx0.997$. Note that reduction of $g/\gamma$ below $0.9$ leads to disappearance of bistability and solitons. It should be mentioned that the $\boldsymbol{\Psi_-}$ family that is always unstable in the $\mathcal{PT}$-symmetric coupler becomes stable in the forced system.

Figures \ref{psiplus}(a,b) show the amplitudes of flat solutions vs $h$ at $\delta=1$ and imbalance $g/\gamma=1$ and $0.9$, respectively. For this positive $\delta$ value both  $\boldsymbol\Psi_-$  and  $\boldsymbol\Psi_+$ soliton families can coexist. However, in this case the existence of localized soliton states is limited to the low values of $h\lesssim0.02$ as shown in Figs. \ref{psiplus}(c,d). All $\boldsymbol\Psi_-$ solitons in Fig.\ref{psiplus}(c) are unstable while the $\boldsymbol\Psi_+$ are stable within a tunable and finite range of the driving $h$, as shown in Fig.\ref{psiplus}(d). The latter figure includes stability prediction for the $\boldsymbol\Psi_+$ family in the $\mathcal{PT}$-symmetric case with pump. We emphasize that in that particular case, such prediction is based on a clearly observed Hopf threshold. However, because stability is always marginal for $g=\gamma$, arbitrarily small instabilities cannot be detected numerically and require specific analysis \cite{BarashPRA2013}.

\begin{figure}
\centering
\includegraphics[width=.5\textwidth]{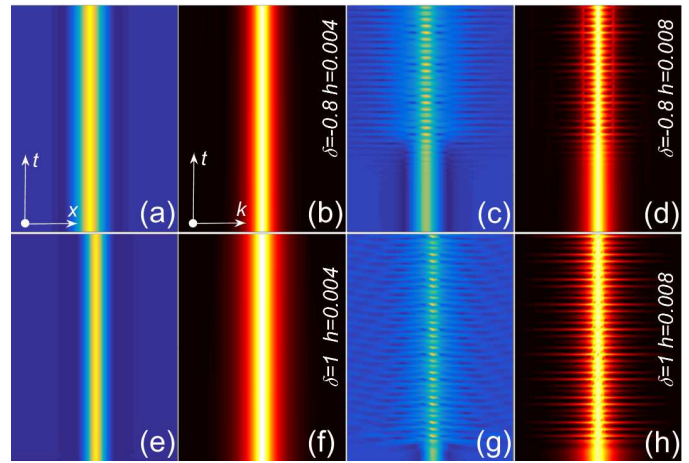}
\caption{(a,b,e,f) stable vs (c,d,g,h) unstable propagations along $4000$ roundtrips in direct (a,c,e,g) and Fourier (b,d,f,h) spaces of $\boldsymbol{\Psi_-}$ solitons for $\delta=-0.8$ (a-d), and $\boldsymbol{\Psi_+}$ solitons for $\delta=1$ (e-h). Stable (unstable) propagations correspond to $h=0.004$ ($h=0.008$). $\gamma=0.5$, $g/\gamma=0.995$ (0.95) for $\boldsymbol{\Psi_{-(+)}}$ and $h_H\approx0.0046$ (0.0058). $x\in[-30, 30]$ and $\mathrm{k}\in[-10, 10]$.\label{proppsi}}
\end{figure}

Figure \ref{proppsi} shows examples of soliton propagation verifying the stability predictions made in Figs. \ref{psimin} and \ref{psiplus}. Simulations were initiated with the numerically computed solitons from both families. In agreement with predictions, solitons found for $h$ values below the Hopf thresholds, $h_H$, exhibit a perfectly stable dynamics (Fig. \ref{proppsi}(a,b,e,f)), while those found for $h>h_H$ transform into breathers  (Fig. \ref{proppsi}(c,d,g,h)), that in the final run after many roundtrips may decay into flat states.

In summary, we have shown that two stable cavity soliton families exist in driven microring dimers with gain and loss when a small imbalance between gain and loss is introduced. Each of the soliton families is associated with nonlinear resonances that are well-separated in detuning as long as the system operates far from the symmetry breaking point. Consequently, the stability regions for each family are attained for different and tunable intervals of the cavity detuning. Our results can be generalized to more complex microring arrays.

{\bf Funding.} Severo Ochoa program (SEV-2015-0522) of the Government of Spain; Fundaci\'{o} Cellex; Fundaci\'{o} Mir-Puig; Generalitat de Catalunya; CERCA; the Leverhulme Trust (RPG-2015-456); ITMO University visiting professorship via the government of Russia grant 074-U01; RFBR (17-02-00081).
%
%
%
%
%

\end{document}